\documentclass[a4paper,UKenglish,cleveref, autoref]{lipics-v2021}
\usepackage{amsmath,amssymb,mathtools,booktabs}
\usepackage{algorithm}
\usepackage[noend]{algpseudocode}
\usepackage{tikz}
\usetikzlibrary{arrows.meta,positioning,calc}
\nolinenumbers

\hideLIPIcs

\title{Internal Queries for Maximal Closed Substrings}
\titlerunning{Internal Queries for Maximal Closed Substrings}

\author{Simone {Faro}}{Department of Mathematics and Computer Science, University of Catania
Viale Andrea Doria 6, 95125 Catania, Italy \and \url{https://www.dmi.unict.it/faro/}}{simone.faro@unict.it}{0000-0001-5937-5796}{}

\author{Francesco Pio Marino}{Department of Mathematics and Computer Science, University of Catania Viale Andrea Doria 6, 95125 Catania, Italy \and \url{https://marfra99x.github.io}}
{francesco.marino@phd.unict.it}{0000-0003-4722-9542}{}

\author{Arianna Pavone}{Department of Mathematics and Computer Science, University of Palermo Via Archirafi 34, 90123 Palermo, Italy \and \url{https://www.ariannapavone.com}}
{ariannamaria.pavone@unipa.it}{0000-0002-8840-6157}{}

\authorrunning{S. Faro, F. P. Marino, and A. Pavone}
\Copyright{Simone Faro, Francesco Pio Marino, and Arianna Pavone}

\EventEditors{}
\EventLongTitle{38th Annual Symposium on Combinatorial Pattern Matching (CPM 2027)}
\EventShortTitle{CPM 2027}
\EventAcronym{CPM}
\EventYear{2027}
\EventDate{June 16--18, 2027}
\EventLocation{London, United Kingdom}
\SeriesVolume{0}
\ArticleNo{0}
\newcommand{\Closed}{\mathsf{Closed}}
\newcommand{\calR}{\mathcal R}
\newcommand{\calL}{\mathcal L}
\newcommand{\calM}{\mathcal M}
\newcommand{\Next}{\operatorname{next}}
\newcommand{\rank}{\operatorname{rank}}
\newcommand{\select}{\operatorname{select}}
\newcommand{\LCA}{\operatorname{LCA}}
\newcommand{\LCE}{\operatorname{LCE}}
\newcommand{\off}{\operatorname{off}}
\newcommand{\depth}{\operatorname{depth}}
\newcommand{\one}{\mathbf 1}
\newcommand{\polylog}{\operatorname{polylog}}
\begin{document}
\maketitle

\begin{abstract}
Searching for repetitions inside a selected part of a text raises a subtle
boundary problem: removing neighbouring characters can create new maximal answers.
We study this effect for maximal closed substrings (MCSs). A string is closed if
it is a single letter or has a nonempty proper prefix occurring exactly twice, as
its prefix and suffix; a closed occurrence is maximal if neither available
one-character extension is closed. Previous work established $O(n\log n)$ bounds
on the global MCS families of a length-$n$ text and developed enumeration
algorithms. Internal queries instead ask for the MCSs of a chosen substring, with
maximality evaluated inside that substring. Separating boundary occurrences from
interior ones gives an $O(n\log n)$-word index that reports the $c$ answers in
$O(\log n+c)$ time and supports counting and longest queries in $O(\log n)$ time.
Our main result reduces reporting space to $O(n)$ words, with
$O((c+1)\log^{1+\varepsilon} n)$ query time and $O(n\log^{1+\varepsilon} n)$
construction time for any fixed $\varepsilon>0$. Thus, after preprocessing,
reporting requires no additive scan of the queried substring, which is
particularly useful for long queries with few answers. Independently, we prove
that at most $3(s-1)$ MCS occurrences of any length-$s$ substring cross a fixed
gap between consecutive characters, with maximality evaluated inside that
substring. The same bound holds for global MCSs contained in a length-$s$ block. A
family over two letters attains $\lfloor(s-1)/2\rfloor$ crossings for every
$s\ge3$, showing that the linear order of this local density bound is optimal.
\end{abstract}
\clearpage
\section{Introduction}
\label{sec:intro}

A substring, or \emph{factor}, is a contiguous part of a text. Restricting
attention to one factor can change which repetitions are maximal, because
its boundaries remove possible extensions. We study this effect for closed strings.
A \emph{border} is a nonempty proper prefix that is also a suffix; a
string is \emph{closed} if it is a single letter or has a border occurring
only at its two ends, possibly with overlap. A maximal closed substring
(MCS) is a closed occurrence with no closed one-character extension on
either side \cite{BadkobehEtAl2022,BadkobehEtAl2026}.

An internal query $[i,j]$ treats $T[i..j]$ as an independent string and
reports, counts, or selects a longest MCS occurrence. Intervals count
separately even when their string values agree. In $T=\mathtt{aab}$,
$[2,2]$ is not an MCS because its left extension $\mathtt{aa}$ is closed,
but it becomes an MCS of $T[2..3]=\mathtt{ab}$. Filtering the text's
global MCSs therefore misses answers created at query boundaries.

Call a closed occurrence \emph{right-maximal} or \emph{left-maximal}
when it has no closed one-character extension on that side. Although
there can be quadratically many closed occurrences, these one-sided
families have tight size $O(n\log n)$ \cite{Kosolobov2024}.
Our starting point splits the answers into four disjoint families:
global MCSs strictly inside the query, right-maximal closed prefixes,
left-maximal closed suffixes, and the whole query if closed.
This gives an explicit index for all three operations. Our main result
reduces reporting space to $O(n)$ words while keeping query time dependent
on the output size, rather than adding a scan of the queried factor.
\Cref{tab:results} states the bounds; space means final index space.
Such an index is particularly useful when many long factors are queried
but only a few occurrences need to be returned. For example, every
nonempty factor of $\mathtt a^n$ has just one MCS: the whole factor.

\begin{table}[ht]
\centering
\caption{Internal-MCS indices on a text of length $n$; $c$ is the number of
reported occurrences; $\varepsilon>0$ is any fixed constant. The compact construction uses $O(n\log n)$ temporary
words in addition to its $O(n)$-word final index.}
\label{tab:results}
\begin{tabular}{@{}llll@{}}
\toprule
Index & Space (words) & Query time & Construction time\\
\midrule
Explicit & $O(n\log n)$ & Reporting: $O(\log n+c)$ & $O(n\log^2 n)$\\
 & & Counting / longest: $O(\log n)$ &\\
Compact & $O(n)$ & Reporting: $O((c+1)\log^{1+\varepsilon} n)$ & $O(n\log^{1+\varepsilon} n)$\\
\bottomrule
\end{tabular}
\end{table}

The compact index describes each candidate by a \emph{record}. Records
with a common start, their \emph{anchor}, form an ordered list. We can
continue the enumeration from any saved state, so we keep a restart state,
or \emph{checkpoint}, every $\Theta(\log n)$ records. Only $O(n)$
checkpoints remain; missing records are reconstructed on demand.

To select which records to reconstruct, order global MCSs by their left
and then right endpoints. A range minimum query (RMQ) returns the position
of a smallest value in an array interval. A compact encoding answers these
queries on right endpoints without retaining the endpoint array itself.
The left boundary selects a suffix of the order; its minimum right
endpoint either supplies an answer or rules out the whole range.
Repeating this test selects only $O(c+1)$ records for reconstruction,
where $c$ is the number of answers.
The encoding chooses which records to reconstruct; checkpoints determine how.

Independently, we bound how many global MCSs contained in a length-$s$ block
can cross a fixed gap between consecutive characters: at most $3(s-1)$.
The bound also holds with maximality evaluated inside the block, and its
linear order is optimal even over two letters. This describes local
repetition density; it is not an ingredient of the index.

\subsection{Related work}
Closed words arise from studying intervals between consecutive copies
of a pattern \cite{Durand1998,CarpiDeLuca2001,Fici2011,Fici2017}.
There can be quadratically many distinct closed factors
\cite{BadkobehFiciLiptak2015,ParshinaPuzynina2024}. Algorithms address
partitioning a text into closed factors \cite{BadkobehEtAl2016}, counting
distinct closed factors \cite{MienoEtAl2025}, and enumerating MCSs
\cite{BadkobehEtAl2022,BadkobehEtAl2026}. Kosolobov's \emph{closed repeats}
correspond to non-singleton MCSs through the consecutive copies of their
longest borders, which cannot be extended together on either side.
The one-sided versions give tight $O(n\log n)$ population bounds and
optimal global enumeration \cite{Kosolobov2024}. Jain and Mhaskar give compact
representations and further algorithms \cite{JainMhaskar2025,JainMhaskar2026};
we reuse their partition of closed lengths. Processing a growing text
\cite{ShibataEtAl2026} or a text encoded by lengths of equal-letter blocks
\cite{UmezakiEtAl2026} addresses different input regimes.

Fischer and K\"oppl~\cite{FischerKoppl2026} enumerate the $k$ global MCSs
in $O(n\sqrt{\log n}+k\log^{\varepsilon}n)$ time and $O(n)$ words.
Their ordered candidates and next-occurrence/common-prefix jumps already
provide the local enumeration rules used here; we do not claim these rules
as new. \Cref{sec:anchors} gives the precise correspondences.

Internal pattern matching locates a chosen pattern within a factor
\cite{KociumakaEtAl2015,KociumakaEtAl2024}. Other internal queries find
short strings absent from a factor \cite{BadkobehEtAl2022ISAW}, match
patterns from a dictionary \cite{ChenEtAl2026}, or study coverage by
possibly overlapping copies of a shorter string \cite{CrochemoreEtAl2026}.
Consecutive-occurrence queries \cite{AkramMieno2025} supply the witnessing
pattern; our queries supply only endpoints, so each answer's border must
be recovered. Kosolobov's substring queries likewise do not report or
count MCSs with query-relative maximality. Our contribution is a retained
$O(n)$-word interface for record access by rank and endpoint-threshold
searches at arbitrary anchors after preprocessing. The index combines this
interface with the four boundary cases and one RMQ directory, which selects
only $O(c+1)$ global records for reconstruction.
The explicit index is a baseline using standard range-searching structures;
the cut bound is a separate combinatorial contribution.

\section{Preliminaries}
\label{sec:prelim}

Let $T[1..n]$ be a text over an integer alphabet whose symbols can be
sorted in $O(n)$ time, and let
$T^R$ denote its reversal. We use a word random-access machine (RAM) with word size
$\Theta(\log n)$ bits. All logarithmic bounds are understood with
$\max\{2,n\}$; constant-size texts can be handled directly. An
occurrence is an interval $[a,b]$, distinct from its string value $T[a..b]$.
We write $\one[P]$ for one when a statement $P$ is true and zero otherwise.

Closedness depends only on the occurrence itself. Maximality additionally
tests its two neighbours, but treats a missing extension as automatically
maximal. We make this boundary convention explicit below.

\begin{definition}
A string is \emph{closed} if it is a single letter or has a border occurring
only as its prefix and suffix. Otherwise it is \emph{open}. Write
$\Closed(a,b)$ when $T[a..b]$ is closed. A closed occurrence $[a,b]$ is
\emph{right-maximal} if $b=n$ or $T[a..b+1]$ is open, and
\emph{left-maximal} if $a=1$ or $T[a-1..b]$ is open. An MCS is maximal
on both sides.
\end{definition}

Closedness is intrinsic to the string, and can equivalently be witnessed
by its longest border. In $\mathtt{ababa}$ the longest border
$\mathtt{aba}$ has exactly two overlapping occurrences, so the string is
closed. Maximality here concerns \emph{one-character} extensions:
$\mathtt a$ is a right-maximal closed prefix of $\mathtt{aba}$ even
though the longer prefix $\mathtt{aba}$ is closed too. We include
singletons, assigning them longest-border length zero.

Let $\calR_T,\calL_T$ be the right- and left-maximal closed families,
and $\calM_T=\calR_T\cap\calL_T$. Each has size $O(n\log n)$
\cite{Kosolobov2024}. Let $\calM_T(i,j)$ be the MCS occurrences of
the independent string $T[i..j]$, expressed in text coordinates; put
$c=|\calM_T(i,j)|$. Reporting returns this set, counting returns $c$,
and longest returns one of its maximum-length intervals. A longest closed
occurrence in a query is necessarily an MCS: an available closed extension
would contradict its maximum length. Thus longest-MCS and longest-closed
queries have the same optimum and witnesses, whereas their reporting and
counting versions are different.

The indices use three standard building blocks. First, a bitvector is an
array of zeros and ones. On a bitvector $B$, $\rank_\sigma(B,t)$ counts bits equal to $\sigma\in\{0,1\}$
through position $t$, with rank zero at $t=0$, and
$\select_\sigma(B,h)$ locates the $h$-th such bit. These operations
take constant time in $|B|+o(|B|)$ bits, a \emph{succinct} representation
\cite{RamanRamanSatti2007}; all list ranks are one-based.

Second, the suffix tree of $T\$$ groups suffixes by shared prefixes,
where $\$$ is a unique terminator. Root-to-leaf paths spell the suffixes;
shared prefixes use the same path, and chains without branching are
contracted into string-labelled edges. Its $O(n)$ explicit nodes are the
edge endpoints; a position strictly within an edge is implicit. The
\emph{path label} of a node is the concatenation of edge labels from
the root, and its \emph{string depth} is that label's length. The
\emph{locus} of a factor is the explicit or implicit point spelling it
on a path from the root. The leaf
$z_a$ represents the suffix starting at $a$. The lowest common ancestor (LCA)
$\LCA(z_a,z_q)$ is the deepest node shared by their root paths. Its
string depth equals the longest common extension (LCE), $\LCE(a,q)$, the
length of the longest common prefix of those suffixes. LCA queries are
constant-time with linear-word support. A
level-ancestor query returns the ancestor at a specified \emph{edge
depth}, the number of edges from the root, and has the same bounds
\cite{BenderFarachColton2004}. It identifies the child of an ancestor
toward a specified leaf. The suffix array lists suffix starts in
lexicographic order; a node's descendant leaves form one contiguous
interval in this array. A range-successor query returns the smallest
value exceeding a given threshold in such an interval.
For the compact index we use the linear-word range-successor
structure of \cite[Proposition~9.7(a)]{KociumakaEtAl2024}, based on
\cite{BelazzouguiPuglisi2016}: for any fixed $\varepsilon>0$, it has
$O(\log^{\varepsilon}n)$ query time and $O(n\sqrt{\log n})$
construction time. Throughout the compact-index bounds, $\varepsilon>0$
is fixed independently of $n$, and hidden constants may depend on it.
Its input is the suffix-array permutation of $[1,n]$; the
terminator-only suffix is omitted. Every queried subtree interval is
stored explicitly.

Third, an RMQ encoding on an array of length $m$ uses $2m+o(m)$ bits
and returns the position of a minimum in constant time, without needing
the array after construction \cite{FischerHeun2011}. Retrieving the
value at that position is a separate operation. This distinction lets
us store only an encoding and reconstruct selected endpoints when needed.
We consistently break ties by array position.

\section{Separating the query boundaries}
\label{sec:boundary}

\begin{figure}[ht]
\centering
\begin{tikzpicture}[x=0.57cm,y=0.47cm,>=Stealth,font=\small]
\draw[gray] (0,0)--(15,0);
\draw[dashed] (2,-0.25)--(2,4.4) node[above] {$i$};
\draw[dashed] (13,-0.25)--(13,4.4) node[above] {$j$};
\foreach \l/\r/\y/\lab in {5/10/3.5/interior global MCS,2/8/2.5/right-maximal prefix,7/13/1.5/left-maximal suffix,2/13/0.5/whole query if closed}{
\draw[very thick,blue!60!black] (\l,\y)--(\r,\y);
\draw (\l,\y-0.12)--(\l,\y+0.12);
\draw (\r,\y-0.12)--(\r,\y+0.12);
\node[anchor=west] at (15.5,\y) {\lab};}
\end{tikzpicture}
\caption{The four answer families are distinguished by contact with the two
query boundaries. Only the interior family requires two endpoint constraints.}
\label{fig:boundary}
\end{figure}
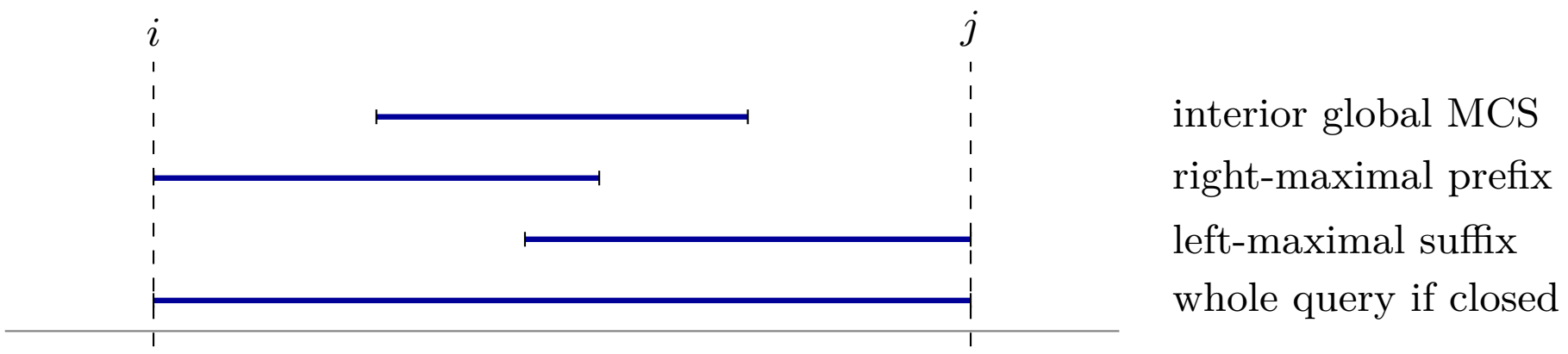

Restricting the text can change maximality only where an extension
disappears. Classify a closed occurrence by whether it touches neither
query boundary, just the left, just the right, or both. An interior
occurrence must pass both global maximality tests. At a boundary, the
missing test can be dropped; for the whole query, only closedness remains.
This gives the following decomposition, where $\dot\cup$ denotes
disjoint union (see \cref{fig:boundary}).

\begin{lemma}[Boundary decomposition]
\label{lem:boundary}
For $1\le i\le j\le n$,
\begin{align*}
\calM_T(i,j)={}&\{[a,b]\in\calM_T:i<a\le b<j\}\\
&\mathbin{\dot\cup}\{[i,b]\in\calR_T:b<j\}
\mathbin{\dot\cup}\{[a,j]\in\calL_T:i<a\}\\
&\mathbin{\dot\cup}
\begin{cases}\{[i,j]\},&\Closed(i,j),\\\varnothing,&\text{otherwise}.\end{cases}
\end{align*}
\end{lemma}
\begin{proof}
An occurrence strictly inside the query retains both one-character
extensions, so local and global maximality agree. A prefix touching only
the left boundary has automatic left maximality, and its right extension
is unchanged; hence it must be globally right-maximal. The suffix case is
symmetric. The whole query has neither extension and contributes exactly
when closed. These arguments prove both inclusions. The four possibilities
for boundary contact are mutually exclusive.
\end{proof}

\section{A fast explicit index}
\label{sec:explicit}

The explicit index stores all one-sided candidates. With one endpoint
fixed, a boundary query becomes a threshold search in an ordered list;
only the interior family needs a two-dimensional search. To also handle
the whole query, the lists must reveal which intervening lengths are
closed, not just which lengths are maximal.

For each start $a$, let $K_a$ be its number of right-maximal occurrences.
Represent its $h$-th occurrence, for $1\le h\le K_a$, by
$(\rho_{a,h},\beta_{a,h})$, where $\rho$ is its length and $\beta$
its longest-border length. The represented interval is $[a,a+\rho-1]$.
Order these records by increasing length; their total size is
$O(n\log n)$.
Mark each possible length by one if the corresponding prefix is closed
and zero otherwise. Every maximal run of ones ends at a right-maximal
record. It therefore suffices to recover where each run begins; the
border lengths of successive records supply that start. Specifically, put
\begin{equation}
\label{eq:classes}
\pi_{a,1}=1,\qquad
\pi_{a,h}=\rho_{a,h}-\beta_{a,h}+\beta_{a,h-1}+1\quad(h>1).
\end{equation}
Then $\Closed(a,a+d-1)$ holds exactly when
$d\in\bigcup_h[\pi_{a,h},\rho_{a,h}]$. This known partition follows
from the open--closed prefix identity
\cite{BadkobehEtAl2022,DeLucaFiciZamboni2017,JainMhaskar2025};
\cref{app:classes} recalls the argument. Binary search on these ordered,
disjoint classes tests closedness in $O(\log n)$ time. The reversed text
supplies the symmetric left-maximal lists.

We next filter for maximality on the other side. Prepending a character
to a closed string preserves closedness exactly when its two longest-border
copies have the same preceding character, or left context. Comparing them is
therefore enough to accept or reject a right-maximal record. For a record
$[a,b]$ with border length $\beta$,
\begin{equation}
\label{eq:flag}
[a,b]\in\calM_T\quad\Longleftrightarrow\quad
a=1\ \text{or}\ T[a-1]\ne T[b-\beta].
\end{equation}
Indeed, for a non-singleton closed string, its left extension is closed
exactly when the characters preceding the two copies of its longest border
agree. The singleton case is the equality test on a length-two string.
\Cref{app:extension} proves the criterion, including exclusion of a
different shorter witness for the extension.

Map each global MCS $[a,b]$ to $(a,b)$ with weight $b-a+1$.
For the interior family the query region is
\begin{equation}
\label{eq:orthant}
D(i,j)=[i+1,n]\times[1,j-1].
\end{equation}
Since $a\le b$, this two-constraint region, or \emph{orthant}, selects
exactly the interior occurrences. Standard geometric structures report
its $k$ points in $O(\log n+k)$ time \cite{McCreight1985} or count them
in $O(\log n)$ \cite{Navarro2014}. Selecting a maximum-weight point
among $N$ points takes $O(\log N)$ time with $O(N)$ words and
$O(N\log N)$ construction. \Cref{app:geometry} proves these bounds
by dividing query space into regions with the same answer, using
\cite[Lemma~5]{FarzanMunroRaman2012}.
Combining the interior result with the two boundary lists and the
whole-query test now resolves all three operations.

\begin{theorem}[Explicit index]
\label{thm:explicit}
An $O(n\log n)$-word index, constructible in $O(n\log^2 n)$ time,
supports reporting in $O(\log n+c)$ time and counting and longest-MCS
queries in $O(\log n)$ time.
\end{theorem}
\begin{proof}
Enumerate the one-sided records using the $O(n\log n)$-time construction
of \cite{Kosolobov2024}, add singletons, filter global MCSs by
\eqref{eq:flag}, and sort the anchored lists. Build the classes and
the three geometric structures above. There are $N=O(n\log n)$ records,
so sorting and geometric preprocessing cost $O(N\log N)$ time and
$O(N)$ words.

For reporting, query $D(i,j)$, report all right-maximal records anchored
at $i$ with endpoint below $j$, all left-maximal records ending at $j$
with start above $i$, and $[i,j]$ if closed. Binary search finds the
selected prefix or suffix of each ordered list. For counting, add the
orthant count, the two selected list sizes, and $\one[\Closed(i,j)]$.
For longest, return the whole query if closed; otherwise compare the
maximum-weight interior point, the longest admissible prefix, and the
longest admissible suffix, ignoring empty families. These two boundary
maxima are the last admissible right endpoint and first admissible left
endpoint. Correctness follows from \cref{lem:boundary}; its disjointness
also avoids duplicate suppression. All searches cost $O(\log n)$,
and reporting additionally scans exactly the output.
\end{proof}

The threshold variant, asking whether an MCS of length at least $k$ exists,
has the same $O(\log n)$ bound by comparing the longest answer with $k$.

\section{Compact access through suffix-tree breakpoints}
\label{sec:anchors}

To avoid storing every endpoint, we reconstruct short stretches of a
list from checkpoints. This first requires enumeration without scanning
all possible lengths. Let $\calR(a)$ be the right-maximal list at $a$
and $\calR^+(a)$ its non-singleton sublist, ordered by increasing right
endpoint. We will support access by rank in $O(n)$ total words.

For an explicit suffix-tree ancestor $x$ of $z_a$, let
$\operatorname{Occ}(x)$ be the suffix starts in $[1,n]$ below $x$,
that is, the occurrence starts of its path label. Define
\[
\Next_a(x)=\min(\operatorname{Occ}(x)\cap[a+1,n]),
\]
with value $+\infty$ if there is no later occurrence. The leaves of $x$
form a suffix-array interval, so a range-successor query evaluates this
value in $O(\log^{\varepsilon} n)$ time. Descending the path to $z_a$ only removes
occurrences: $\Next_a$ is nondecreasing. We call a maximal consecutive
segment with the same finite value a \emph{plateau}.
A \emph{breakpoint} is a non-root explicit node $x$ with finite
$\Next_a(x)$ whose child $y$ toward $z_a$ has a larger value.

Here a context is a neighbouring character, with the terminator used at
the right text boundary. Two consecutive copies of a prefix delimit a
closed occurrence. When their next characters differ, extending that
occurrence to the right destroys closedness. Breakpoints capture exactly
these stopping positions. This reformulates the candidate correspondence
of \cite[Lemma~4 and Section~5]{FischerKoppl2026} using explicit tree nodes.

\begin{lemma}[Records and breakpoints]
\label{lem:breakpoints}
The breakpoints for anchor $a$ correspond bijectively, in order, to
$\calR^+(a)$. A breakpoint of string depth $t$ and successor $q$
represents $[a,q+t-1]$, with longest-border length $t$.
\end{lemma}
\begin{proof}
Let $U$ be the path label of a breakpoint $x$, with $|U|=t$.
Its copies at $a$ and $q$ are consecutive, since $q=\Next_a(x)$,
and have different following contexts: $q$ lies below $x$ but outside
its child toward $a$. In $T[a..q+t-1]$, any occurrence of $U$ must
start between $a$ and $q$, so these are its only two copies and the
interval is closed. A longer border of length $\gamma>t$ would
start its suffix copy at $q+t-\gamma$, strictly between $a$ and $q$;
its first $t$ letters would give a third copy of $U$. Thus $t$ is
exactly the longest-border length.
Different following contexts make the interval right-maximal, by the
right-hand version of the extension criterion. This is the right
closed-repeat correspondence of \cite{Kosolobov2024}.

Conversely, the longest border of a non-singleton right-maximal closed
interval has consecutive occurrences at $a$ and $q$, with different
following contexts. Its locus is consequently an explicit branching node,
and $q$ disappears in the child toward $a$, giving the unique breakpoint spelling that border. The
terminator supplies the differing context when the second copy ends at
$n$. Consecutive breakpoints have strictly increasing depth and successor
position, hence strictly increasing represented right endpoint.
\end{proof}

After the next occurrence changes, its new copy stays available as long
as the two suffixes share a prefix. Their lowest common ancestor marks
where this agreement ends, so we can jump there directly. This is the
next-occurrence/common-prefix step of \cite[Section~4, procedure \texttt{fill}]{FischerKoppl2026}.

\begin{lemma}[Direct plateau continuation]
\label{lem:lca}
Let $x$ be a breakpoint at $a$, and $y$ its child toward $z_a$. Put
$q'=\Next_a(y)$. If $q'=+\infty$, there is no later breakpoint.
Otherwise the next breakpoint is
\[
x'=\LCA(z_a,z_{q'}).
\]
It can be found in $O(\log^{\varepsilon} n)$ time.
\end{lemma}
\begin{proof}
For finite $q'$, both leaves lie below $y$, so $x'$ is a descendant of
$y$, possibly $y$ itself. At every node from $y$ through $x'$, the leaf
$q'$ remains present. No position in $(a,q')$ occurred below $y$, and
descending cannot add one. Thus the successor is exactly $q'$ throughout
this path segment. In the child of $x'$ toward $z_a$, the leaf $q'$
disappears, so the successor increases. Hence $x'$ is the end of this
plateau and the next breakpoint; no node between $y$ and $x'$
can be an intervening breakpoint, since its successor has not changed. The infinite case follows from subtree
containment. The operations are one range successor, a child lookup, and
one constant-time LCA.
\end{proof}

Initialization uses the same argument: take $y$ to be the root child
toward $z_a$, compute $q=\Next_a(y)$, and, if finite, start with
$\LCA(z_a,z_q)$. If $q$ is infinite, $\calR^+(a)$ is empty.
The resulting enumeration visits every non-singleton record once and
costs $O(\log^{\varepsilon} n)$ per record and per anchor.

A schematic of this continuation appears in \cref{fig:lca}.

We can now trade stored records for short reconstruction walks. Keeping
one checkpoint every logarithmic number of records divides the
$O(n\log n)$ population into only $O(n)$ blocks. Access by rank decodes
at most one block. Endpoint thresholds and closedness tests also locate
one block by searching its saved endpoint or class start. A threshold
rank marks where the ordered endpoints cross a supplied bound; it
determines the size and longest member of the selected prefix or suffix.

\begin{theorem}[Checkpointed anchor access]
\label{thm:anchors}
There is an $O(n)$-word representation of the one-sided lists, constructible
in $O(n\log^{1+\varepsilon} n)$ time, supporting list sizes in $O(1)$ time, access to
any ranked record in $O(\log^{1+\varepsilon} n)$ time, and $q$ consecutive records in
$O(\log^{1+\varepsilon} n+(q+1)\log^{\varepsilon} n)$ time. It also tests $\Closed(i,j)$ and
finds a boundary-family threshold rank in $O(\log^{1+\varepsilon} n)$ time.
\end{theorem}
\begin{proof}
Set $L=\lceil\log n\rceil$ and partition each $\calR^+(a)$ into
blocks of at most $L$ consecutive records. Store the first record of each
block as a checkpoint, containing $(a,x,q)$, its represented length and
border length, and its closed-length class start from \eqref{eq:classes}.
These are constant-word restart states for \cref{lem:lca}. There are
\[
\sum_a\left\lceil |\calR^+(a)|/L\right\rceil
\le \sum_a|\calR^+(a)|/L+n=O(n)
\]
checkpoints. Encode a group of size $d$ as $1^d0$, where $1^d$ means
$d$ consecutive one-bits; concatenating these groups gives a unary
delimiter bitvector. For example, sizes $(2,0,1)$ give $110010$.
Rank/select then identifies a group and a rank within it. Apply this
encoding to list lengths and checkpoint counts, using $O(n\log n)$
bits. A bitvector $F$ with one bit per non-singleton
logical record marks the global MCSs using \eqref{eq:flag}. Two length-$n$
bitvectors mark right-maximal and globally maximal singletons; a possible
right-maximal singleton is the first entry of the full list $\calR(a)$.

Rank/select locates the block of a requested local rank in constant time.
At most $L-1$ continuations recover its record, costing $O(\log^{1+\varepsilon} n)$.
Once decoding starts, each additional record costs $O(\log^{\varepsilon} n)$.
For a threshold on increasing right endpoints, binary-search the stored
checkpoint endpoints and decode at most one block to obtain the exact
rank. The search costs $O(\log n)$ and decoding $O(\log^{1+\varepsilon} n)$.

For closedness, answer length one directly. For a longer query at anchor
$i$, binary-search the stored class starts for the last checkpoint start
not exceeding $d=j-i+1$. Only that block can contain a class covering
$d$: all preceding blocks end before its first class, and all later blocks
start after $d$. Decode this block and test membership in its disjoint
classes. Each subsequent class start is obtained from the previous border
length using \eqref{eq:classes}; the first is stored at the checkpoint.
If a singleton is a separate first class it cannot contain $d>1$.
If no singleton is right-maximal, the first non-singleton record has the
initial class starting at one. Both cases are therefore covered.

Build the same index on $T^R$ for the left-maximal lists. Reversal
preserves closedness and exchanges sides, but reverses the required
increasing-start order: original rank $h$ corresponds to rank $K-h+1$
in a reversed list of size $K$. Random access and thresholds have the
same bounds. For sequential output in increasing-start order, process
reversed checkpoint blocks from right to left, decoding each forward and
buffering at most $L$ records before reversing its output. This costs
$O((L+q)\log^{\varepsilon} n)$ time and $O(L)$ temporary words.

The two suffix-tree indices, LCA/level-ancestor support, and range-successor structures
occupy $O(n)$ words; checkpoints and all bitvectors fit the same bound.
Enumerating $O(n\log n)$ logical records at $O(\log^{\varepsilon} n)$ each, plus
$O(n\sqrt{\log n})$ range-successor preprocessing, takes
$O(n\log^{1+\varepsilon} n)$ time. This also covers the linear-time
construction of the trees and bitvector support.
\end{proof}

For either query boundary, its selected prefix or suffix of an anchored
list can consequently be counted or maximised in $O(\log^{1+\varepsilon} n)$ time,
or reported with $O(\log^{1+\varepsilon} n+(k+1)\log^{\varepsilon} n)$ cost for $k$ records.
This handles all three non-interior terms of \cref{lem:boundary}.
\Cref{app:decoder} specifies the decoder and full query interface,
including reversal, empty lists, partial blocks, and singleton offsets.

\section{Linear-space reporting with one global RMQ}
\label{sec:report}

We now report global MCSs whose endpoints lie in $[L,R]$, where
$1\le L\le R\le n$. The two conditions are $a\ge L$ and $b\le R$;
such a pair of endpoint inequalities is a \emph{dominance query}.
Sorting by $a$ makes its first condition easy: the eligible records form
a suffix of the order. For the second condition we need only find a
smallest $b$, test it, and repeat on the remaining ranges. This is why
an encoding of comparisons can replace an explicit array of coordinates.

Concatenate $\calR^+(1),\ldots,\calR^+(n)$, with total length
$N=O(n\log n)$, and put $K_a^+=|\calR^+(a)|$. The anchor representation
already contains the delimiter and global-MCS flag vectors
\[
D=1^{K_1^+}0\,1^{K_2^+}0\cdots1^{K_n^+}0,\qquad F[1..N].
\]
Filter this concatenation by the one-bits of $F$, obtaining the logical
sequence $G[1..M]$ of all non-singleton global MCSs, sorted by $(a,b)$.
The order in $G$ is conceptual: the flags identify its members without
storing another list. Let $\off(a)$ count all unfiltered records before
anchor $a$. The zero delimiters locate where that anchor's group begins:
\begin{equation}
\label{eq:offset}
\off(1)=0,\qquad
\off(a)=\select_0(D,a-1)-(a-1)=\sum_{x<a}K_x^+\quad(a>1).
\end{equation}

Decoding first undoes the filtering, then identifies the anchor group
and the position within it. Global rank counts one-bits of $F$; local
rank counts unfiltered records in one non-singleton list. Construction
produces right endpoints in this same filtered order, so the RMQ and
decoder name the same records without an additional coordinate map.

\begin{lemma}[Global rank decoding]
\label{lem:global-rank}
For $1\le h\le M$, the record $G[h]$ can be reconstructed in
$O(\log^{1+\varepsilon} n)$ time with no further stored coordinate sequence.
\end{lemma}
\begin{proof}
Let $p=\select_1(F,h)$, its rank before filtering. Put
\[
s=\select_1(D,p),\qquad
a=\rank_0(D,s)+1,\qquad g=p-\off(a).
\]
The one-bit of $D$ at $s$ belongs to its $a$-th group, and $g$ is its
rank within that group. Thus $G[h]$ is exactly $\calR^+(a)[g]$.
All translations take constant time; \cref{thm:anchors} decodes the record
in $O(\log^{1+\varepsilon} n)$ time. Empty anchor groups are handled by their zero
delimiters and do not alter the formulas.
\end{proof}

During construction form the temporary sequence $A[h]=b_h$ for
$G[h]=[a_h,b_h]$ and build its succinct RMQ encoding. The retained
encoding occupies $2M+o(M)$ bits. It returns a rank attaining a minimum
right endpoint, and \cref{lem:global-rank} recovers the value when needed;
the array $A$ itself is discarded. We use the following conventions throughout: if $M=0$ no RMQ is
built; an empty rank range returns before any select, decoding, or RMQ
call; empty anchor groups contribute only a zero to $D$; RMQ ties
choose the smaller global rank. Singletons are excluded from $D,F,G,A$
and handled by their separate bitvector.

For a query $[L,R]$, count the flagged records anchored before $L$ and
skip them. The records with $a_h\ge L$ form precisely the
suffix beginning at
\begin{equation}
\label{eq:first-rank}
h_0=\rank_1(F,\off(L))+1.
\end{equation}
This follows because $\off(L)$ counts all non-singleton one-sided records
anchored strictly before $L$, and the flags select the global ones.
The remaining constraint is $b_h\le R$.
For example, $T=\mathtt{abacaba}$ has non-singleton global records
$[1,3],[1,7],[3,5],[5,7]$. Containment in $[3,5]$ retains the last two
by start: minimum endpoint five reports $[3,5]$, while seven is rejected.
This gives the interior non-singleton answer $\mathtt{aca}$ for query
$[2,6]$; its whole string $\mathtt{bacab}$ is also locally maximal.
\Cref{app:example} gives the complete decoding example.

Within this suffix, a minimum endpoint is a certificate for the whole
current range. If it exceeds $R$, every record in that range can be
discarded. Otherwise it supplies one answer; removing it leaves two
smaller ranges to search. Start \cref{alg:global} on $[h_0,M]$.
Each RMQ returns
a rank, not an endpoint value: the global-rank decoder is invoked only
for that selected record. The endpoint array $A$ in the notation denotes
the array used at construction time, not a retained query-time array.

\begin{algorithm}[H]
\caption{Global non-singleton reporting, initially called with $(h_0,M,R)$}
\label{alg:global}
\begin{algorithmic}[1]
\Procedure{Explore}{$\ell,r,R$}
  \If{$\ell>r$} \State \Return \EndIf
  \State $h\gets\operatorname{RMQ}_A(\ell,r)$
  \State $[a,b]\gets\operatorname{DecodeGlobal}(h)$
  \If{$b>R$} \State \Return \EndIf
  \State \Call{Report}{$[a,b]$}
  \State \Call{Explore}{$\ell,h-1,R$}
  \State \Call{Explore}{$h+1,r,R$}
\EndProcedure
\end{algorithmic}
\end{algorithm}

\begin{lemma}[Implicit dominance reporting]
\label{lem:global-report}
The procedure reports exactly the non-singleton global MCSs contained
in $[L,R]$ in $O((k+1)\log^{1+\varepsilon} n)$ time, where $k$ is their number.
Its directory occupies $O(n\log n)$ bits in addition to anchor access.
\end{lemma}
\begin{proof}
Every record considered satisfies $a\ge L$, by \eqref{eq:first-rank},
and the invariant persists under recursive subdivision. If the minimum
$b$ exceeds $R$, the entire subrange violates the remaining constraint.
Otherwise the reported interval satisfies $L\le a\le b\le R$.
The two recursive ranges partition all remaining records, proving by
induction that every feasible record is output exactly once. An upper
constraint on $a$ is unnecessary, since $a\le b\le R$ for each output.

Each successful nonempty call reports a different record and generates
two children. With $k$ successes there are at most $k+1$ unsuccessful
leaves, including empty ranges. Thus at most $2k+1$ calls decode a
record, even when $k=0$. The RMQs and translations are constant-time,
and each decoding costs $O(\log^{1+\varepsilon} n)$. The sole additional sequence
encoding takes $2M+o(M)=O(n\log n)$ bits. If the selected suffix is
empty, the procedure returns without accessing the RMQ.
\end{proof}

The pseudocode describes the logical recursion. To bound its working
stack, recursively visit the smaller of the two rank ranges and process
the larger by tail iteration. Every nested call then halves its range,
so $O(\log(M+1))=O(\log n)$ temporary words suffice. The output
order is unrestricted, and the same records and pruning tests are used.

The pieces now fit the four boundary cases: checkpoint access handles
the two boundary lists, minimum queries handle the interior, and a
closedness test handles the whole query. The families are disjoint, so
their costs add according to the total output, while records remain implicit.

\begin{theorem}[Linear-space internal reporting]
\label{thm:compact}
For every fixed $\varepsilon>0$, a text of length $n$ admits an $O(n)$-word index reporting
$\calM_T(i,j)$ in $O((c+1)\log^{1+\varepsilon} n)$ time. It can be constructed in
$O(n\log^{1+\varepsilon} n)$ time using $O(n\log n)$ temporary words.
\end{theorem}
\begin{proof}
For the strict interior use $L=i+1$, $R=j-1$. If $L>R$, this family
is empty. Otherwise apply \cref{lem:global-report}, and report the global
singletons in $[L,R]$ using their separate bitvector. The two boundary
families and whole-query closedness use \cref{thm:anchors}. The families
are disjoint by \cref{lem:boundary}, so correctness follows and their
output sizes sum to $c$. The interior costs $O((k+1)\log^{1+\varepsilon} n)$;
singleton reporting is linear in its output and each boundary costs
$O(\log^{1+\varepsilon} n+(k'+1)\log^{\varepsilon} n)$. Together these fit the claimed time.

The anchor representation, flags, delimiters, singleton vectors, and
RMQ encoding use $O(n\log n)$ bits, equivalent to $O(n)$ words.
For construction, including range-successor preprocessing, enumerate the
anchor records and their flags in
$O(n\log^{1+\varepsilon} n)$ time. During enumeration append the right endpoint of
each flagged non-singleton record to a temporary array, in anchor order.
This array is exactly $A$. Build its RMQ encoding in $O(M)$ time and
discard it. The temporary record streams and array occupy
$O(n\log n)$ words. No owner assignment, second sorting of global
records, or persistent endpoint array is required.
The range-successor preprocessing also fits this workspace:
its underlying construction uses $O(n\log n)$ working bits,
with auxiliary lookup tables fitting in $O(n)$ words
\cite{BelazzouguiPuglisi2016}.
The bitvector and RMQ constructions use at most $O(N+M+n)$ temporary
words. Hence the total construction workspace is $O(n\log n)$ words.
\end{proof}

For a query of length $s$, fresh enumeration using
\cite{FischerKoppl2026} costs
$O(s\sqrt{\log s}+c\log^{\varepsilon}s)$, whereas our query bound
has no input-length term. After preprocessing, our bound is
$o(s\sqrt{\log s})$ whenever
$(c+1)\log^{1+\varepsilon}n=o(s\sqrt{\log s})$.
This describes a useful regime of long, small-output queries; it is not
an improvement for every query, and a single query need not amortise
index construction. Compact counting and longest queries remain open.

The output factor is a reconstruction cost: RMQ navigation performs only
$O(c+1)$ selections, but each selected record can require
$\Theta(\log n)$ successor queries. The stronger
$O(\polylog n+c)$ time bound, with $\polylog n$ denoting a fixed
power of $\log n$, would require, within this reconstruction scheme, a faster or batched decoder.
Exact counting and longest queries in linear space are not supplied by
this directory: minima of $b$ identify reportable records, but do not give
their number or the maximum weight $b-a+1$ under both endpoint constraints.

\section{A linear bound across a cut}
\label{sec:cut}

Let $B=[u,v]$ have length $s\ge2$ and fix a cut after $c$, where
$u\le c<v$. An interval crosses the cut if $a\le c<b$.
The following result concerns global MCSs fully contained in $B$ and
is independent of \cref{thm:compact}. Related local density questions concern substrings that occur only
once but lose uniqueness on shortening \cite{FujimaruMienoInenaga2025}.

For a non-singleton MCS $[a,b]$, let its longest border have length $t$
and put $q=b-t+1$. Its \emph{arms} are $[a,a+t-1]$ and $[q,b]$;
they may overlap. An arm $[x,y]$ crosses the cut exactly when
$x\le c<y$.

We distinguish cuts through a border copy from cuts between the two
copies. For the first case, a separation property lets us split the
right arms of MCSs with a common starting position into two disjoint
collections: odd-numbered arms and even-numbered arms. Each collection can cross a cut at most
once. Reversing the text gives the corresponding statement for left arms.

\begin{lemma}[Arm packing]
\label{lem:packing}
For each fixed start $a$, at most two right arms of MCSs starting at
$a$ cross a fixed cut. For each fixed end $b$, at most two left arms
of MCSs ending at $b$ cross it.
\end{lemma}
\begin{proof}
List the right closed repeats at $a$ by increasing length
$t_1<\cdots<t_k$ and let $q_x$ be their next occurrences.
Kosolobov's separation lemma gives $q_{x+2}>q_x+t_x$
\cite[Lemma~2]{Kosolobov2024}. Therefore their right arms are disjoint
within each parity of $x$. A cut crosses at most one arm in each parity,
hence at most two. MCSs form a subfamily. Reversal proves the other side.
\end{proof}

If neither arm crosses, the two copies lie on opposite sides of the
cut. For any fixed border value, consecutiveness then leaves at most one
possible pair: the last copy on the left and the first on the right.
To count these border values, we use \emph{maximal repeats}: repeated
factors with at least two different preceding contexts and two different
following contexts, treating missing contexts as boundary sentinels.
This is maximality of a repeated \emph{value} across its occurrences,
distinct from maximality of a closed \emph{interval} under extension.

\begin{lemma}[Gap injection]
\label{lem:gap}
Among MCSs crossing the cut whose arms do not cross it, the map to
the string value of the longest border is injective. Every image is a
nonempty maximal repeat of the independent string $T[u..v]$.
\end{lemma}
\begin{proof}
Such an interval satisfies $a+t-1\le c<q$: its border copies lie on
opposite sides and are consecutive occurrences. Maximality of the MCS
means these copies have distinct preceding contexts and distinct following
contexts, by the extension criterion. On restriction to $B$, the following
context of the left copy and preceding context of the right copy stay
inside $B$. Either other context stays unchanged or becomes a boundary
sentinel, distinct from alphabet characters. The inequalities survive, so
the border value is a maximal repeat of the block.

For a fixed border value $U$, the first copy must be its last occurrence
wholly left of the cut: a later such occurrence would start between
$a$ and $q$. Similarly the second must be its first occurrence wholly
right of the cut. An occurrence crossing the cut and starting between
$a$ and $q$ would also contradict consecutiveness; in that case no
admissible pair exists. Thus there is at most one pair, and $U$
uniquely determines $a,q$ and hence $[a,q+|U|-1]$.
\end{proof}

The two counting arguments now combine: crossing arms contribute at
most $2(s-1)$ occurrences in total, and the remaining occurrences use
at most $s-1$ distinct repeated values.

\begin{theorem}[Linear cut bound]
\label{thm:cut}
At most $3(s-1)$ global MCS occurrences contained in a length-$s$
block cross a fixed internal cut.
\end{theorem}
\begin{proof}
Single letters cross no cut. Partition the others into those whose right
arm crosses, those whose left arm crosses but right arm does not, and
those with neither arm crossing. This priority also handles overlapping
arms. In the first class, $u\le a\le c-1$; by \cref{lem:packing}
there are at most $2(c-u)$ members. The second class has endpoints
$c+1\le b\le v$, giving at most $2(v-c)$ members.
By \cref{lem:gap}, the third class injects into distinct nonempty maximal
repeats of $T[u..v]$. Each is a non-root internal suffix-tree node, and
a length-$s$ string with a unique terminator has at most $s-1$ such
nodes. Summing gives $2(c-u)+2(v-c)+(s-1)=3(s-1)$.
\end{proof}

The same statement applies to query-relative maximality by treating
the queried factor as the text.

\begin{corollary}[Local cut bound]
\label{cor:local-cut}
For a factor $T[i..j]$ of length $s\ge2$, at most $3(s-1)$ of its
MCS occurrences cross any fixed internal cut, with maximality evaluated
inside the factor.
\end{corollary}
\begin{proof}
Apply \cref{thm:cut} to the independent text $T[i..j]$, taking
its whole length-$s$ interval as the block.
\end{proof}

A single $\mathtt b$ between two blocks of $\mathtt a$, equal in length
or differing by one, explains why a sublinear bound is impossible. For example, $\mathtt{aaabaaa}$ has
nested MCS occurrences $\mathtt{aba}$, $\mathtt{aabaa}$, and
$\mathtt{aaabaaa}$ crossing the cut before $\mathtt b$.
Extending just one side, when possible, creates a third copy of every
possible border. Nesting is compatible with one-character maximality.

\begin{proposition}[Tightness of the linear order]
\label{prop:cut-lower}
For every integer $s\ge3$, there is a binary string of length $s$ with at least
$\lfloor(s-1)/2\rfloor$ MCSs crossing one cut.
\end{proposition}
\begin{proof}
Put $m=\lfloor(s-1)/2\rfloor$. For odd $s$, take
$T=\mathtt a^m\mathtt b\mathtt a^m$; for even $s$, take
$T=\mathtt a^m\mathtt b\mathtt a^{m+1}$. Fix the cut immediately
before the unique $\mathtt b$. For $1\le k\le m$, the interval
$[m-k+1,m+k+1]$ spells $\mathtt a^k\mathtt b\mathtt a^k$.
Its border $\mathtt a^k$ occurs exactly twice, so it is closed.
Every available one-character extension has blocks of $\mathtt a$
of lengths $k$ and $k+1$. Any nonempty border of such an extension is
$\mathtt a^r$ with $r\le k$: a proper border containing the unique
$\mathtt b$ would force two occurrences of that letter. Each possible
border has $(k-r+1)+(k+1-r+1)\ge3$ occurrences, so the extension is open.
For $k=m$ the left extension is absent; the right extension is absent
for odd $s$ and is open by the same argument for even $s$.
Thus all $m$ intervals are MCSs and cross the same cut.
\end{proof}

Consequently the worst-case cut density is $\Theta(s)$, even on a binary
alphabet. This does not determine the optimal constant in the upper bound.

\section{Conclusions}
\label{sec:conclusion}

Checkpointed access to one-sided families and a single global RMQ
encoding give linear-word internal reporting; the cut density is
$\Theta(s)$ even on binary texts. Open questions include
$O(\polylog n+c)$ reporting, polylogarithmic counting and longest queries
in linear space, and construction with linear workspace. An
$O(n\log n)$-bit index for global containment aggregation would give
internal counting or longest queries with only
$O(\log^{1+\varepsilon}n)$ additional boundary cost
(\cref{app:aggregation}). The optimal constant in the cut bound remains
unknown.

\section*{Use of generative artificial intelligence}
Generative AI tools, including OpenAI Codex, were used for exploratory
analysis, verification support, and manuscript preparation. Exploratory
uses included discussing proof strategies, examining alternative
data-structure choices, and developing candidate examples and extremal
constructions. Verification support included checking intermediate
arguments and complexity calculations against the cited literature,
identifying potential boundary cases, and drafting and running programs
for exhaustive checks on small instances. Manuscript preparation included
assistance with English expression, organisation, notation, and LaTeX
editing.

AI-generated suggestions were not regarded as authoritative sources.
The authors retain full responsibility for the mathematical claims,
proofs, references, and final content of the manuscript.

\label{end:main}
\clearpage
\bibliography{references}

@article{Durand1998,
  author  = {Durand, Fabien},
  title   = {A Characterization of Substitutive Sequences Using Return Words},
  journal = {Discrete Mathematics},
  volume  = {179},
  number  = {1--3},
  pages   = {89--101},
  year    = {1998},
  doi     = {10.1016/S0012-365X(97)00029-0}
}

@article{CarpiDeLuca2001,
  author  = {Carpi, Arturo and de Luca, Aldo},
  title   = {Periodic-Like Words, Periodicity, and Boxes},
  journal = {Acta Informatica},
  volume  = {37},
  number  = {8},
  pages   = {597--618},
  year    = {2001},
  doi     = {10.1007/PL00013314}
}

@inproceedings{Fici2011,
  author    = {Fici, Gabriele},
  title     = {A Classification of Trapezoidal Words},
  booktitle = {Proceedings of the 8th International Conference on Words},
  series    = {Electronic Proceedings in Theoretical Computer Science},
  volume    = {63},
  pages     = {129--137},
  year      = {2011},
  doi       = {10.4204/EPTCS.63.18}
}

@article{DeLucaFiciZamboni2017,
  author  = {De Luca, Alessandro and Fici, Gabriele and Zamboni, Luca Q.},
  title   = {The Sequence of Open and Closed Prefixes of a {Sturmian} Word},
  journal = {Advances in Applied Mathematics},
  volume  = {90},
  pages   = {27--45},
  year    = {2017},
  doi     = {10.1016/j.aam.2017.04.007}
}

@article{Fici2017,
  author  = {Fici, Gabriele},
  title   = {Open and Closed Words},
  journal = {Bulletin of the EATCS},
  volume  = {123},
  pages   = {140--149},
  year    = {2017},
  url     = {https://bulletin.eatcs.org/index.php/beatcs/article/view/508}
}

@inproceedings{BadkobehFiciLiptak2015,
  author    = {Badkobeh, Golnaz and Fici, Gabriele and Lipt{\'a}k, Zsuzsanna},
  title     = {On the Number of Closed Factors in a Word},
  booktitle = {Language and Automata Theory and Applications},
  series    = {Lecture Notes in Computer Science},
  volume    = {8977},
  pages     = {381--390},
  year      = {2015},
  publisher = {Springer International Publishing},
  doi       = {10.1007/978-3-319-15579-1_29}
}

@article{BadkobehEtAl2016,
  author  = {Badkobeh, Golnaz and Bannai, Hideo and Goto, Keisuke and I, Tomohiro and Iliopoulos, Costas S. and Inenaga, Shunsuke and Puglisi, Simon J. and Sugimoto, Shiho},
  title   = {Closed Factorization},
  journal = {Discrete Applied Mathematics},
  volume  = {212},
  pages   = {23--29},
  year    = {2016},
  doi     = {10.1016/j.dam.2016.04.009}
}

@inproceedings{BadkobehEtAl2022,
  author    = {Badkobeh, Golnaz and De Luca, Alessandro and Fici, Gabriele and Puglisi, Simon J.},
  title     = {Maximal Closed Substrings},
  booktitle = {String Processing and Information Retrieval},
  series    = {Lecture Notes in Computer Science},
  volume    = {13617},
  pages     = {16--23},
  year      = {2022},
  publisher = {Springer},
  doi       = {10.1007/978-3-031-20643-6_2}
}

@article{BadkobehEtAl2026,
  author  = {Badkobeh, Golnaz and De Luca, Alessandro and Fici, Gabriele and Puglisi, Simon J.},
  title   = {Finding Maximal Closed Substrings},
  journal = {Theoretical Computer Science},
  volume  = {1060},
  pages   = {115628},
  year    = {2026},
  doi     = {10.1016/j.tcs.2025.115628}
}

@article{Kosolobov2024,
  author  = {Kosolobov, Dmitry},
  title   = {Closed Repeats},
  journal = {CoRR},
  volume  = {abs/2410.00209},
  year    = {2024},
  doi     = {10.48550/arXiv.2410.00209},
  url     = {https://arxiv.org/abs/2410.00209}
}

@article{ParshinaPuzynina2024,
  author  = {Parshina, Olga G. and Puzynina, Svetlana},
  title   = {Finite and Infinite Closed-Rich Words},
  journal = {Theoretical Computer Science},
  volume  = {984},
  pages   = {114315},
  year    = {2024},
  doi     = {10.1016/j.tcs.2023.114315}
}

@inproceedings{JainMhaskar2025,
  author    = {Jain, Samkith K. and Mhaskar, Neerja},
  title     = {Efficient Computation of Closed Substrings},
  booktitle = {String Processing and Information Retrieval},
  series    = {Lecture Notes in Computer Science},
  volume    = {16073},
  pages     = {172--187},
  year      = {2025},
  publisher = {Springer Nature Switzerland},
  doi       = {10.1007/978-3-032-05228-5_15}
}

@article{JainMhaskar2026,
  author  = {Jain, Samkith K. and Mhaskar, Neerja},
  title   = {Efficient Algorithms to Compute Closed Substrings},
  journal = {Theory of Computing Systems},
  volume  = {70},
  number  = {3},
  pages   = {48},
  year    = {2026},
  doi     = {10.1007/s00224-026-10290-x},
  url     = {https://doi.org/10.1007/s00224-026-10290-x}
}

@inproceedings{MienoEtAl2025,
  author    = {Mieno, Takuya and Takahashi, Shun and Seto, Kazuhisa and Horiyama, Takashi},
  title     = {Online and Offline Algorithms for Counting Distinct Closed Factors via Sliding Suffix Trees},
  booktitle = {SOFSEM 2025: Theory and Practice of Computer Science},
  series    = {Lecture Notes in Computer Science},
  volume    = {15539},
  pages     = {172--183},
  year      = {2025},
  publisher = {Springer Nature Switzerland},
  doi       = {10.1007/978-3-031-82697-9_13}
}

@inproceedings{ShibataEtAl2026,
  author    = {Shibata, Hiroki and Umezaki, Haruki and Mieno, Takuya and Nakashima, Yuto and Inenaga, Shunsuke},
  title     = {Online Computation of Maximal Closed Substrings},
  booktitle = {String Processing and Information Retrieval ({SPIRE} 2026)},
  year      = {2026},
  publisher = {Springer},
  note      = {To appear; preprint available as arXiv:2607.00612},
  url       = {https://arxiv.org/abs/2607.00612}
}

@inproceedings{UmezakiEtAl2026,
  author    = {Umezaki, Haruki and Shibata, Hiroki and Nakashima, Yuto and Inenaga, Shunsuke},
  title     = {Compact Enumeration of Maximal Closed Substrings in Run-Length Encoded Strings},
  booktitle = {String Processing and Information Retrieval ({SPIRE} 2026)},
  year      = {2026},
  publisher = {Springer},
  note      = {To appear; preprint available as arXiv:2607.02838},
  url       = {https://arxiv.org/abs/2607.02838}
}

@inproceedings{FujimaruMienoInenaga2025,
  author    = {Fujimaru, Hiroto and Mieno, Takuya and Inenaga, Shunsuke},
  title     = {On the Number of {MUS}s Crossing a Position},
  booktitle = {String Processing and Information Retrieval},
  series    = {Lecture Notes in Computer Science},
  volume    = {16073},
  pages     = {124--132},
  year      = {2025},
  publisher = {Springer Nature Switzerland},
  doi       = {10.1007/978-3-032-05228-5_11}
}

@inproceedings{KociumakaEtAl2015,
  author    = {Kociumaka, Tomasz and Radoszewski, Jakub and Rytter, Wojciech and Wale{\'n}, Tomasz},
  title     = {Internal Pattern Matching Queries in a Text and Applications},
  booktitle = {Proceedings of the Twenty-Sixth Annual ACM--SIAM Symposium on Discrete Algorithms},
  pages     = {532--551},
  year      = {2015},
  publisher = {SIAM},
  doi       = {10.1137/1.9781611973730.36}
}

@article{KociumakaEtAl2024,
  author  = {Kociumaka, Tomasz and Radoszewski, Jakub and Rytter, Wojciech and Wale{\'n}, Tomasz},
  title   = {Internal Pattern Matching Queries in a Text and Applications},
  journal = {SIAM Journal on Computing},
  volume  = {53},
  number  = {5},
  pages   = {1524--1577},
  year    = {2024},
  doi     = {10.1137/23M1567618}
}

@article{CrochemoreEtAl2026,
  author  = {Crochemore, Maxime and Iliopoulos, Costas S. and Radoszewski, Jakub and Rytter, Wojciech and Straszy{\'n}ski, Juliusz and Wale{\'n}, Tomasz and Zuba, Wiktor},
  title   = {Internal Quasiperiod Queries},
  journal = {Theoretical Computer Science},
  volume  = {1066},
  pages   = {115747},
  year    = {2026},
  doi     = {10.1016/j.tcs.2026.115747}
}

@inproceedings{AkramMieno2025,
  author    = {Akram, Waseem and Mieno, Takuya},
  title     = {Sorted Consecutive Occurrence Queries in Substrings},
  booktitle = {36th Annual Symposium on Combinatorial Pattern Matching (CPM 2025)},
  series    = {Leibniz International Proceedings in Informatics},
  volume    = {331},
  pages     = {24:1--24:15},
  year      = {2025},
  publisher = {Schloss Dagstuhl -- Leibniz-Zentrum f{\"u}r Informatik},
  doi       = {10.4230/LIPIcs.CPM.2025.24}
}

@article{BadkobehEtAl2022ISAW,
  author  = {Badkobeh, Golnaz and Charalampopoulos, Panagiotis and Kosolobov, Dmitry and Pissis, Solon P.},
  title   = {Internal Shortest Absent Word Queries in Constant Time and Linear Space},
  journal = {Theoretical Computer Science},
  volume  = {922},
  pages   = {271--282},
  year    = {2022},
  doi     = {10.1016/j.tcs.2022.04.029}
}

@inproceedings{ChenEtAl2026,
  author    = {Chen, Jingbang and Dai, Jiangqi and Mang, Qiuyang and Shi, Qingyu and Xu, Tingqiang},
  title     = {Nearly Optimal Internal Dictionary Matching},
  booktitle = {34th Annual European Symposium on Algorithms ({ESA} 2026)},
  series    = {Leibniz International Proceedings in Informatics},
  volume    = {388},
  pages     = {58:1--58:22},
  year      = {2026},
  publisher = {Schloss Dagstuhl -- Leibniz-Zentrum f{\"u}r Informatik},
  doi       = {10.4230/LIPIcs.ESA.2026.58}
}

@article{McCreight1985,
  author  = {McCreight, Edward M.},
  title   = {Priority Search Trees},
  journal = {SIAM Journal on Computing},
  volume  = {14},
  number  = {2},
  pages   = {257--276},
  year    = {1985},
  doi     = {10.1137/0214021}
}

@article{Navarro2014,
  author  = {Navarro, Gonzalo},
  title   = {Wavelet Trees for All},
  journal = {Journal of Discrete Algorithms},
  volume  = {25},
  pages   = {2--20},
  year    = {2014},
  doi     = {10.1016/j.jda.2013.07.004}
}

@article{BenderFarachColton2004,
  author  = {Bender, Michael A. and Farach-Colton, Mart{\'i}n},
  title   = {The Level Ancestor Problem Simplified},
  journal = {Theoretical Computer Science},
  volume  = {321},
  number  = {1},
  pages   = {5--12},
  year    = {2004},
  doi     = {10.1016/j.tcs.2003.05.002}
}

@article{RamanRamanSatti2007,
  author  = {Raman, Rajeev and Raman, Venkatesh and Satti, Srinivasa Rao},
  title   = {Succinct Indexable Dictionaries with Applications to Encoding $k$-ary Trees, Prefix Sums and Multisets},
  journal = {ACM Transactions on Algorithms},
  volume  = {3},
  number  = {4},
  pages   = {43:1--43:25},
  year    = {2007},
  doi     = {10.1145/1290672.1290680}
}

@inproceedings{FarzanMunroRaman2012,
  author    = {Farzan, Arash and Munro, J. Ian and Raman, Rajeev},
  title     = {Succinct Indices for Range Queries with Applications to Orthogonal Range Maxima},
  booktitle = {Automata, Languages, and Programming},
  series    = {Lecture Notes in Computer Science},
  volume    = {7391},
  pages     = {327--338},
  year      = {2012},
  publisher = {Springer},
  doi       = {10.1007/978-3-642-31594-7_28}
}

@article{FischerHeun2011,
  author  = {Fischer, Johannes and Heun, Volker},
  title   = {Space-Efficient Preprocessing Schemes for Range Minimum Queries on Static Arrays},
  journal = {SIAM Journal on Computing},
  volume  = {40},
  number  = {2},
  pages   = {465--492},
  year    = {2011},
  doi     = {10.1137/090779759}
}

@article{Kirkpatrick1983,
  author       = {David G. Kirkpatrick},
  title        = {Optimal Search in Planar Subdivisions},
  journal      = {{SIAM} J. Comput.},
  volume       = {12},
  number       = {1},
  pages        = {28--35},
  year         = {1983},
  url          = {https://doi.org/10.1137/0212002},
  doi          = {10.1137/0212002},
  bibsource    = {dblp computer science bibliography, https://dblp.org}
}

@article{FischerKoppl2026,
 author = {Fischer, Johannes and K{\"o}ppl, Dominik},
 title = {Output-Sensitive Enumeration of Maximal Closed Substrings in Linear Space},
 journal = {CoRR}, volume = {abs/2610.38114}, year = {2026},
 doi = {10.48550/arXiv.2610.38114}, url = {https://arxiv.org/abs/2610.38114}
}

@inproceedings{BelazzouguiPuglisi2016,
 author = {Belazzougui, Djamal and Puglisi, Simon J.},
 title = {Range Predecessor and {Lempel-Ziv} Parsing},
 booktitle = {Proceedings of the Twenty-Seventh Annual ACM--SIAM Symposium on Discrete Algorithms},
 pages = {2053--2071}, year = {2016}, publisher = {SIAM},
 doi = {10.1137/1.9781611974331.ch143}
}

\clearpage
\appendix
\section{Closed-length classes}
\label{app:classes}

Each class is a consecutive stretch of closed prefix lengths. Counting
closed prefixes cumulatively lets us recover a class size by subtraction;
its endpoint then determines its start. For $X[1..m]$, let $P(d)$ be
the greatest length of a nonempty prefix
occurring at least twice in $X[1..d]$, or zero if none exists. The
open--closed prefix identity gives
\begin{equation}
\label{eq:oc}
P(d)=\sum_{r=1}^d\one[X[1..r]\text{ is closed}]-1.
\end{equation}
At a closed prefix of length $d$, $P(d)$ is its longest-border length.
These facts hold for arbitrary finite words
\cite[Lemmas~2 and~3]{BadkobehEtAl2022}, referring to
\cite[Remark~8]{DeLucaFiciZamboni2017}.

Right-maximal lengths $\rho_h$ are exactly the endpoints of maximal
runs of ones in the open--closed prefix sequence. Put
$\beta_h=P(\rho_h)$. The first run starts at one. For $h>1$,
subtracting \eqref{eq:oc} at consecutive endpoints gives
$\beta_h-\beta_{h-1}$ entries, hence start
$\pi_h=\rho_h-\beta_h+\beta_{h-1}+1$. This proves
\eqref{eq:classes}, including class order and disjointness.

\section{The one-character extension criterion}
\label{app:extension}

Equal preceding characters extend both border copies together. For the
converse, a shorter proposed border would have an extra copy obtained by
shifting it along the repetition already present in the string. Formally,
a positive integer $p$ is a \emph{period} of $X$ if $X[r]=X[r+p]$ whenever both positions
belong to $X$. A border of length $|X|-p$ gives precisely this equality.
The shift by $p$ will rule out any alternative witness for the extension.

\begin{lemma}
Let $X$ be a non-singleton closed string of length $m$, with longest
border $U$ of length $\beta$, and put $p=m-\beta$. For a character
$c$, the string $cX$ is closed if and only if $c=X[p]$.
\end{lemma}
\begin{proof}
If $c=X[p]$, the two copies of $U$ extend left to copies of $cU$,
at the ends of $cX$. An internal copy of $cU$ would yield an additional
copy of $U$ in $X$, contradicting closedness. Hence $cU$ witnesses
closedness of $cX$.

Conversely, let $V$ be a border of $Y=cX$ with no internal occurrence,
and put $\gamma=|V|$ and $r=m+1-\gamma$. Deleting the first
character of the two copies of $V$ shows that $V[2..\gamma]$ is
a border of $X$, possibly empty. Hence $\gamma-1\le\beta$,
or $r\ge p$. The border $U$ makes $p$ a period of $X$.
If $r>p$, shift the suffix copy of $V$ in $Y$ left by $p$.
Its shifted interval is $[r-p+1,m+1-p]$, entirely inside the copy
of $X$ in $Y$, since $r-p+1\ge2$. Periodicity makes it another
copy of $V$, starting strictly between its prefix and suffix copies.
This is a contradiction. Thus $r=p$ and $\gamma=\beta+1$.
The first character of the suffix copy of $V$ is $X[p]$, while its
prefix starts with $c$, giving the required equality.
\end{proof}

Applied to $X=T[a..b]$, this compares $T[a-1]$ with $T[b-\beta]$.
If $a=1$ no left extension is available. If $X$ is a singleton, its
extension has length two and is closed exactly when its two letters
agree, giving the same formula with $\beta=0$. This proves
\eqref{eq:flag}. Reversing gives the right-extension criterion used
in \cref{lem:breakpoints}.

\section{Two-sided weighted maximum}
\label{app:geometry}

Think of each query as a point whose coordinates are its two thresholds.
Nearby queries often have the same winning input point. We partition this
query plane into regions with a common answer; finding the region containing
a query is called \emph{point location}. Horizontal \emph{influence segments}
encode which candidate wins at each query position: the first segment
above that position identifies the answer. The construction below supplies
the bounds used in \cref{thm:explicit}; its sweep processes points from
left to right. A semi-open segment includes one endpoint and excludes
the other; triangulating a region means dividing it into triangles.

\begin{proposition}[Two-sided weighted maximum]
\label{prop:geometry}
Given $N$ weighted two-dimensional points with word-sized coordinates
and weights, an $O(N)$-word static structure can be constructed in
$O(N\log N)$ time to return a maximum-weight point in a two-sided
orthogonal range in $O(\log N)$ time, or report that it is empty.
Empty and constant-size sets are handled directly.
\end{proposition}
\begin{proof}
Reflections reduce the range to $x\le X$, $y\ge Y$.
Sort each coordinate, replacing duplicate values by consecutive distinct
ranks; translate each query threshold by binary search, including the
entire rank block of an endpoint value. Break weight ties by record rank.
For $x\le X$ take the last rank in the block for $X$ (or the last
rank with value below $X$); for $y\ge Y$ take the first rank with
value at least $Y$. Arbitrary orders inside equal-coordinate blocks do
not split an eligible block. Priorities are distinct pairs of weight and
record rank, so geometric perturbations cannot change the selected
maximum weight. This takes $O(N\log N)$ construction time and $O(\log N)$ per query
and preserves the eligible records and maximum-weight answers.

The influence construction of
\cite[Lemma~5, Section~3.3]{FarzanMunroRaman2012} assigns each point at
most one horizontal semi-open segment, possibly empty. The first segment
hit by an upward ray from $(X,Y)$ identifies the maximum-priority eligible
point; no hit means an empty range. Retain the segment endpoints and
labels explicitly.

The cited construction sweeps points in increasing $x$, keeping active
segments in a balanced tree ordered by height; their priorities increase
from top to bottom. At a point $p$, locate the active segment immediately
above it. If that segment has larger priority, $p$'s influence is empty.
Otherwise delete successive segments below $p$ of smaller priority,
terminating them at $x(p)$, and insert $p$'s segment. A missing segment
above has priority $-\infty$; surviving segments extend to infinity.
Each search, insertion, or deletion takes $O(\log N)$, and each segment
is inserted and deleted at most once. The sweep costs $O(N\log N)$.

Clip the noncrossing segments to a box outside the ranked query positions.
Vertical rays from segment endpoints to the next segment or box boundary
produce an $O(N)$-cell trapezoidal decomposition, constructible by a
standard sweep in $O(N\log N)$ time. Label each cell by the first
horizontal segment above it, or by an empty-range marker, and triangulate
it. Static planar point location \cite{Kirkpatrick1983} then uses $O(N)$
words, linear preprocessing on this triangulated subdivision, and
$O(\log N)$ query time. Place translated thresholds between consecutive
ranks; semi-open conventions settle subdivision boundaries. These bounds,
together with threshold translation, prove the proposition.
\end{proof}

For $N=O(n\log n)$ global MCS points, this gives the longest-query
component and construction bound of \cref{thm:explicit}.

\section{Checkpoint operations and reversal}
\label{app:decoder}
\label{app:reversal}
\label{app:construction}

A checkpoint keeps exactly the state needed to resume a list, so an
access only replays the short portion following that checkpoint. We keep
a possible initial singleton separate from the non-singleton records.
All operations below concern one right-maximal list. Let $\delta_a$
indicate whether $[a,a]$ is right-maximal, so
$K_a=\delta_a+K_a^+$ and non-singleton rank $g$ has full rank
$g+\delta_a$. For $L=\lceil\log n\rceil$, block $j$ starts at
$r_j=1+(j-1)L$ and ends at $e_j=\min\{jL,K_a^+\}$.
Its checkpoint stores $(x,q,\rho,\beta,\pi)$ for rank $r_j$;
the rank itself is implicit. A transient record is
$u=(g,x,q,\rho,\beta,\pi)$, representing $[a,a+\rho-1]$ and
class $[\pi,\rho]$. Its flag is $F[\off(a)+g]$.

The decoder needs only the continuation of \cref{lem:lca}. In the
pseudocode $\depth$ is string depth, and \textsc{Advance} is called
only when the next rank exists, so $q'$ is finite.

\begin{algorithm}[ht]
\caption{Core checkpoint decoder}
\label{alg:decode-block}
\begin{algorithmic}[1]
\Function{Advance}{$a,u$}
  \State $(g,x,q,\rho,\beta,\pi)\gets u$; $y\gets$ child of $x$ toward $z_a$
  \State $q'\gets\Next_a(y)$; $x'\gets\LCA(z_a,z_{q'})$
  \State $\beta'\gets\depth(x')$; $\rho'\gets q'-a+\beta'$
  \State \Return $(g+1,x',q',\rho',\beta',\rho'-\beta'+\beta+1)$
\EndFunction
\Function{AccessPlus}{$a,g$} \Comment{$1\le g\le K_a^+$}
  \State $j\gets1+\lfloor(g-1)/L\rfloor$
  \State $u\gets$ checkpoint $j$, with rank $r_j$
  \While{$u.g<g$} \State $u\gets\Call{Advance}{a,u}$ \EndWhile
  \State \Return $u$
\EndFunction
\end{algorithmic}
\end{algorithm}

The full interface follows from this decoder without further stored data:
\begin{itemize}
\item $\operatorname{AccessRecord}(a,h)$ returns the singleton when
$\delta_a=1$ and $h=1$, and otherwise calls
$\operatorname{AccessPlus}(a,h-\delta_a)$.
\item $\operatorname{FindThreshold}(a,d)$ counts records of length at most
$d$. Binary-search for the last checkpoint $j$ with $\rho\le d$.
If none exists, return $\delta_a\one[d\ge1]$. Otherwise decode ranks
$r_j$ through $e_j$ and return
$\delta_a\one[d\ge1]+r_j-1+\#\{g\in[r_j,e_j]:\rho_g\le d\}$.
Earlier blocks qualify entirely and later blocks do not.
\item $\operatorname{IsClosed}(a,b)$ returns true for $d=b-a+1=1$.
For $d>1$, binary-search for the last checkpoint with $\pi\le d$;
return false if none exists, and otherwise decode that block and test
whether $\pi_g\le d\le\rho_g$ for some record. Disjoint ordered
classes exclude every other block. A separate singleton class contains
only one; if no singleton is right-maximal, the first non-singleton
class begins at one.
\item $\operatorname{Enumerate}(a,h,k)$ returns immediately for $k=0$.
Otherwise emit and remove a requested initial singleton, if present.
If records remain, access the first by its translated non-singleton
rank, emit it, and repeatedly advance and emit until $k$ total records
have been produced. The requested ranks are assumed valid.
\end{itemize}

Empty non-singleton lists have no checkpoints and both predecessor
searches return ``none''. The endpoints $e_j$ handle the final partial
block. Access performs at most $L-1$ continuations; each scanned block
has at most $L$ records and needs no preceding block, since its first
$\pi$ is stored. Thus access, thresholds, and closedness cost
$O(\log^{1+\varepsilon} n)$, and consecutive output costs
$O(\log^{1+\varepsilon} n+(k+1)\log^{\varepsilon} n)$. The checkpoint predecessor searches cost $O(\log n)$, which is
absorbed by $O(\log^{1+\varepsilon}n)$. A state uses $O(1)$ temporary words;
buffering a block uses $O(L)$.

Reversal preserves border length and global maximality. A left-maximal
list ending at $b$ corresponds to the reversed right-maximal list at
$a'=n-b+1$: rank $h$ becomes $K-h+1$, where
$K=|\calL_T(b)|=|\calR_{T^R}(a')|$, and coordinates map as
$[u,v]\mapsto[n-v+1,n-u+1]$. Apply the same formula to non-singleton
sublists with their own size; a singleton is last in the original list
and first in the reversed list. Search thresholds in reversed endpoint
order. For increasing original-start output, visit reversed blocks right
to left, decode each forward, and emit its selected part in reverse.
This visits $O(1+k/L)$ blocks and decodes $O(L+k)$ records, including
the two partial end blocks, with $O(L)$ temporary words.

For the right boundary of $T[i..j]$, put $a'=n-j+1$,
$K=|\calR_{T^R}(a')|$, and
$k_R=\operatorname{FindThreshold}_{T^R}(a',j-i)$.
Indeed, a reversed record of length $\rho$ maps to
$[j-\rho+1,j]$, whose start is greater than $i$ exactly when
$\rho\le j-i$. The selected reversed ranks are $1,\ldots,k_R$;
they correspond to original ranks $K-k_R+1,\ldots,K$.
If $k_R=0$, the family is empty. This formula includes the singleton
through the full-list threshold interface.

The permanent objects and construction workspace are already accounted
for in \cref{thm:anchors,thm:compact}. Neither decoding nor reversal adds
a persistent per-record object: both retain only one transient state or
an $O(L)$-record buffer. Construction writes the RMQ input only in the
original pass; the reversed pass creates its own flags, not a second RMQ.

\section{Residual aggregation}
\label{app:aggregation}

Counts add over disjoint families, while a longest answer is the longest
of their individual answers. The boundary operations already provide both;
only the interior family lacks compact counting and optimisation. Supplying
that component therefore gives the following transfer with explicit costs.

\begin{proposition}[Parametric transfer]
Suppose global MCSs contained in $[L,R]$ can be counted, or one of
maximum length selected, by an $O(n\log n)$-bit structure with
respective query times $q_{\mathrm{cnt}}(n,\ell)$ and
$q_{\mathrm{max}}(n,\ell)$ on intervals of length at most $\ell$.
Together with anchor access and \cref{thm:compact}, this gives an
$O(n)$-word internal index with reporting time
$O((c+1)\log^{1+\varepsilon} n)$ and the respective bounds
\[
O(q_{\mathrm{cnt}}(n,\ell)+\log^{1+\varepsilon} n),\qquad
O(q_{\mathrm{max}}(n,\ell)+\log^{1+\varepsilon} n).
\]
\end{proposition}
\begin{proof}
Apply the assumed structure to $[i+1,j-1]$, or take an empty interior
when this interval is invalid. Counts add over \cref{lem:boundary};
for optimisation, compare the interior maximum and the two boundary
maxima, returning the whole query immediately if closed. All boundary
work takes $O(\log^{1+\varepsilon} n)$ by \cref{thm:anchors}. Singleton global
MCSs may be included in the assumed structure or handled separately.
\end{proof}

\section{A worked example of implicit reporting}
\label{app:example}

For $T=\mathtt{abacaba}$, the non-singleton right-maximal records in
concatenation order are as follows; $F[p]=1$ marks global maximality.

\begin{center}
\begin{tabular}{@{}rrrrrrr@{}}
\toprule
$p$ & Anchor $a$ & Local rank $g$ & Successor $q$ & $\rho$ & $\beta$ & $F[p]$\\
\midrule
1 & 1 & 1 & 3 & 3 & 1 & 1\\
2 & 1 & 2 & 5 & 7 & 3 & 1\\
3 & 2 & 1 & 6 & 6 & 2 & 0\\
4 & 3 & 1 & 5 & 3 & 1 & 1\\
5 & 5 & 1 & 7 & 3 & 1 & 1\\
\bottomrule
\end{tabular}
\end{center}

At anchor one, the breakpoint depths are $1,3$ and successors $3,5$.
After the first breakpoint, the next successor is $5$ and
$\LCA(z_1,z_5)$ has depth three. With illustrative block size two,
only the first record is a checkpoint: one continuation recovers
$(\rho,\beta)=(7,3)$ from $(3,1)$. Their classes are $[3,3]$ and
$[6,7]$, since the full list begins with singleton class $[1,1]$ and
$\pi_2=7-3+1+1=6$. The record $[2,7]$ is filtered out because
its left extension $[1,7]$ is closed. Hence
\[
\begin{gathered}
D=\mathtt{110101001000},\qquad F=\mathtt{11011},\\
G=([1,3],[1,7],[3,5],[5,7]),\qquad A=(3,7,5,7).
\end{gathered}
\]
For global rank $h=3$, decoding gives $p=4$, $s=6$, $a=3$,
$\off(3)=3$, and local rank $g=1$, recovering $[3,5]$.
Only the RMQ encoding of $A$ is retained. The lookup chain is
\[
\begin{gathered}
\underbrace{h=3}_{\text{RMQ rank}}
\xrightarrow{\select_1(F,\cdot)}p=4
\xrightarrow{\select_1(D,\cdot)}s=6
\xrightarrow{\rank_0(D,\cdot)+1}a=3,\\
 a=3,\ p=4
\xrightarrow{p-\off(a)}g=1
\xrightarrow{\text{checkpoint}}[3,5].
\end{gathered}
\]
The values of $A$ shown here explain the encoding; they are not stored
in the final index.

For containment in $[2,6]$, $h_0=3$: RMQ on ranks $[3,4]$ reports
$[3,5]$, then rank four is pruned because its endpoint is $7>6$.
The internal query $T[2..6]=\mathtt{bacab}$ has strict interior
$[3,5]$ and selects the same ranks and non-singleton answer. Its whole
interval is also locally maximal, despite the closed right extension
$[2,7]$ in $T$. Singletons and boundary families use \cref{lem:boundary}.

\section{A schematic of direct continuation}
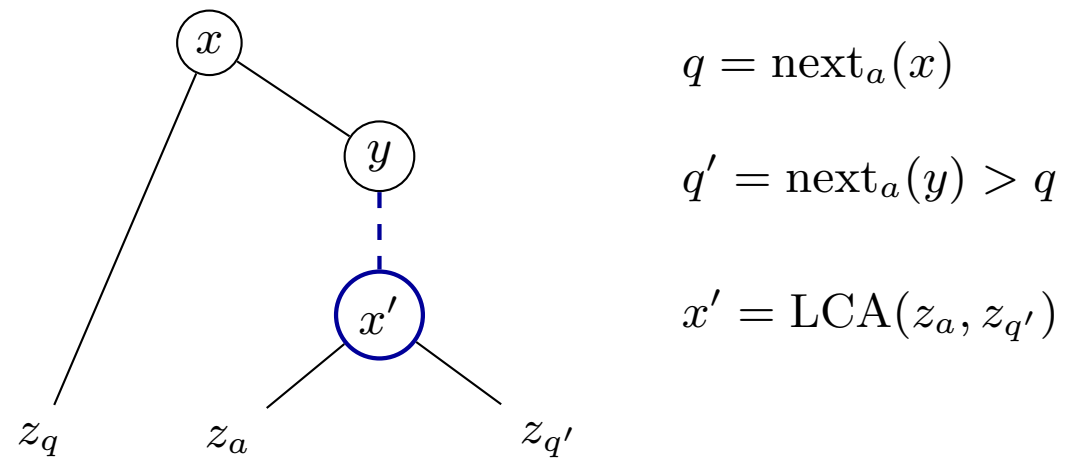
\begin{figure}[H]
\centering
\begin{tikzpicture}[x=1.25cm,y=0.75cm,font=\small,
  vertex/.style={circle,draw,fill=white,inner sep=2pt}]
\node[vertex] (x) at (0,0) {$x$};
\node[vertex] (y) at (0.9,-1) {$y$};
\node[vertex,draw=blue!60!black,thick] (xp) at (0.9,-2.4) {$x'$};
\node (q) at (-0.9,-3.5) {$z_q$};
\node (a) at (0.1,-3.5) {$z_a$};
\node (qp) at (1.8,-3.5) {$z_{q'}$};
\draw (x)--(y);
\draw (x)--(q);
\draw[dashed,thick,blue!60!black] (y)--(xp);
\draw (xp)--(a);
\draw (xp)--(qp);
\node[anchor=west] at (2.4,-0.2) {$q=\Next_a(x)$};
\node[anchor=west] at (2.4,-1.2) {$q'=\Next_a(y)>q$};
\node[anchor=west] at (2.4,-2.4) {$x'=\LCA(z_a,z_{q'})$};
\end{tikzpicture}
\caption{Direct continuation. The old successor $q$ leaves the path at
$x$. The next successor $q'$ remains throughout the dashed path segment,
whose last node is the LCA of the two suffix leaves. Dashed and long
edges may represent paths in the suffix tree.}
\label{fig:lca}
\end{figure}

\end{document}